\documentclass[12pt]{iopart}
\usepackage{iopams}

\usepackage[mathscr]{eucal}
\usepackage{array}
\usepackage{epsfig}
\usepackage{psfrag} 
\usepackage{subfigure}

\usepackage{graphicx}
\usepackage{bm}

\usepackage{pifont}              

\newcommand{\eqref}[1]{(\ref{#1})}

\newcommand{\figref}[1]{figure~\ref{#1}}

\newcommand{\Bb}{{\boldsymbol{\mathnormal b}}}

\newcommand{\Be}{{\boldsymbol{\mathnormal e}}}

\newcommand{\Bg}{{\boldsymbol{\mathnormal g}}}

\newcommand{\Bn}{{\boldsymbol{\mathnormal n}}}

\newcommand{\Bq}{{\boldsymbol{\mathnormal q}}}
\newcommand{\Br}{{\boldsymbol{\mathnormal r}}}
\newcommand{\Bs}{{\pmb{\mathnormal s}}}
\newcommand{\Bt}{{\boldsymbol{\mathnormal t}}}

\newcommand{\Bv}{{\boldsymbol{\mathnormal v}}}

\newcommand{\CC}{{\boldsymbol{\mathnormal C}}}

\newcommand{\BG}{{\boldsymbol{\mathnormal G}}}

\newcommand{\BM}{{\boldsymbol{\mathnormal M}}}

\newcommand{\BT}{{\boldsymbol{\mathnormal T}}}

\newcommand{\superscr}[1]{\ensuremath{{}^{\rm #1}}}

\newcommand{\Balpha }{\ensuremath{\boldsymbol\alpha}}

\newcommand{\Bsigma }{\ensuremath{\boldsymbol\sigma}}
\newcommand{\Brho}{{\boldsymbol{\rho}}}
\newcommand{\Bmu}{{\boldsymbol{\mu}}}

\newcommand{\Beps    }{\ensuremath{\boldsymbol\epsilon}}

\newcommand{\Bepspl }{\ensuremath{\boldsymbol\epsilon\superscr{pl}}}

\newcommand{\Bbetapl}{\ensuremath{\boldsymbol\beta\superscr{pl}}}

\newcommand{\Btau   }{\ensuremath{\boldsymbol\tau}}

\newcommand{\diff}{{\rm d}}

\begin{document}

\title[Continuum Dislocation Dynamics as a Phase Field Theory]{Continuum Dislocation Dynamics as a Phase Field Theory with Conserved Order Parameters}

\author{Yufan Zhang}
\address{Institute for Materials Simulation, University Erlangen-N\"urnberg, Dr. Mack-Strasse 77, 90762 F\"urth, Germany}
\author{Ronghai Wu}
\address{School of Mechanics, Civil Engineering and Architecture, Northwestern Polytechnical University, Xian, 710072, PR China}
\author{Michael Zaiser}
\address{Institute for Materials Simulation, University Erlangen-N\"urnberg, Dr. Mack-Strasse 77, 90762 F\"urth, Germany}
\ead{michael.zaiser@ww.uni-erlangen.de}

\date{\today}


\begin{abstract}
The dynamics of dislocations can be formulated in terms of the evolution of continuous variables representing dislocation densities ('continuum dislocation dynamics'). We show for various variants of this approach that the resulting models can be envisaged in terms of the evolution of order-parameter like variables that strives to minimize a free energy functional which incorporates interface energy-like terms, i.e., as a phase field theory. We show that dislocation density variables obey non-standard conservation laws. These lead, in conjunction with the externally supplied work, to evolution equations that go beyond the classical framework of Ginzburg-Landau vs Cahn-Hilliard equations. The approach is applied to the evolution of dislocation patterns in materials with B1(NaCl) lattice structure.
\end{abstract}

\maketitle

\section{\label{intro} Introduction}

Phase field models consider the evolution of spatially continuous order parameters that evolve to minimize a free energy functional which may include the work of external driving forces. Approaches of this type have been used to model dislocation microstructure evolution in two fundamentally different manners. 

One important class of phase field models for dislocation plasticity are based upon considering the plastic slip magnitudes on  different slip planes as scalar, non conserved order parameters. These models build upon the idea that dislocations constitute boundaries (i.e. linear interfaces) separating regions of different slip on their slip planes. They exploit the capability of phase field models to trace the evolution of (diffuse) interfaces in order to model the evolution of systems of dislocation lines within the phase field formalism. To this end, a Peierls-like energy is introduced which is a periodic function of the slip created by the passage of a single dislocation on a (coarse grained) slip plane. Dislocations then appear naturally as boundaries between domains of different degrees of slip, and their elastic interaction is captured by the elastic energy functional associated with the slip eigenstrains. 

Models of this type were proposed by Koslowski et. al. \cite{koslowski2002phase} for 2D slip in a single plane, and by Wang et. al. \cite{wang2001nanoscale} and Rodney et. al. \cite{rodney2003phase} for 3D dislocation networks, and further developed by Levitas and Javanbakth \cite{levitas2012advanced,javanbakht2016phase} . They allow to resolve the dislocation microstructure on the single-dislocation level and may thus be considered a variant of discrete dislocation dynamics that is particularly suited for describing the co-evolution of dislocation and phase microstructures \cite{levitas2015interaction}. Since the method uses a diffuse interface method to describe the slip discontinuity at the dislcoation core, the core size is effectively controlled by the interface width. This poses problems: Either one chooses the core size to be comparable to the physical core size, which implies a very fine grid resolution and a commensurately high computational cost, or the model will strongly under-estimate the interactions of describing  interfaces with sub-grid resolution \cite{finel2018sharp} may alleviate but cannot fully solve this problem. 

A second class of models operate on a larger scale where dislocations are described not individually but in terms of density-like variables that obey conservation laws, i.e., one is here dealing with conserved order parameters, albeit the nature of the conservation law may differ from that of standard phase field models cast into the framework of the Cahn-Hilliard type equations. Density based phase field modeling was used in the context of dislocations as early as 1970, when Holt \cite{holt1970dislocation} published a study of dislocation patterning that considered the evolution of a scalar density of screw dislocations of the same sign, which was treated as a conserved order parameter -- a study that was clearly crafted after Cahn's seminal work on spinodal decomposition \cite{cahn1961spinodal}. Over the years, several works have appeared which consider the evolution of dislocation densities in a similar spirit. We mention the work of Zaiser et. al. \cite{zaiser2001statistical} who consider patterning of edge dislocations of positive and negative signs whose densities are again treated as conserved order parameters. In that study, in straightforward generalization of Holt's model patterning is considered in absence of an external driving force, while the characteristic length scale of the enmergent patterns was related to generic scaling properties of dislocation systems \cite{zaiser2014scaling}. Variants of this approach were considered in multiple works where dislocation patterning under external loadings was treated in terms of dynamic instabilities associated with a negative density derivative of the mobility function, while still adhering to the formal structure of phase field models with conserved order parameters \cite{groma2016dislocation, wu2018instability, wu2021cell}. Finally, Groma and co-workers considered a phase field model with a conserved density of edge dislocations to describe the  formation and properties of dislocation boundary layers \cite{groma2015scale}. 

Dislocation densities as order parameters can be equally coupled with other order parameters describing the phase microstructure of the material. This was shown by Wu et. al. \cite{wu2017continuum,wu2019effect} in the context of co-evolution of dislocation and precipitate microstructure during creep of Ni base superalloys. 

By considering dislocation densities as scalar, conserved order parameters obeying equations of the Cahn-Hilliard type, density based approaches face the obvious problem that, in plastically deforming crystals, dislocation densities are not in general conserved. Rather, the dislocation density often increases in the course of deformation, giving rise to work hardening, while in special cases (e.g. in nanoscale specimens) it may also decrease, giving rise to exhaustion hardening. Both observations sit uneasily with the concept of dislocation densities as conserved order parameters. However, building a density based model using the Ginzburg-Landau formalism may be even less feasible, since the energy of the dislocation system is a monotonically increasing function of the dislocation density (see the works of Zaiser and Berdichevsky \cite{zaiser2015local,berdichevsky2016energy}), and any treatment of dislocation densities as non conserved order parameters might lead to the erroneous prediction that there are no densities at all. 

To overcome this problem, we start out, in Section 2.1, from a 'hydrodynamic' viewpoint on the Cahn-Hilliard equation which we derive from the standard continuity equation for the hydrodynamic transport of point particles, where the flow velocity is chosen such as to ensure mimimization of the free energy functional. We then transfer the same ideas to Mura's equation for the dislocation density tensor, which can be envisaged as a non standard conservation law describing the transport of a vectorial order parameter. We note that similar ideas were used by Hochrainer within the kinematic formalism of continuum dislocation dynamics \cite{hochrainer2014continuum,hochrainer2015multipole}. Recently, Groma et. al. \cite{groma2021dynamics} applied this formalism to derive transport equations for density variables describing curved dislocations moving on a single slip system. Here we pursue a slightly different approach, starting directly from Mura's equation for the dislocation density tensor which we interpret as a conservation equation. Section 2 explains the general formalism, while in Section 3 we present different applications in the fields of dislocation patterning and for modelling the work hardening contribution of dislocation boundary layers. Section 4 presents conclusions and a brief outlook. 

\section{Conserved and quasi-conserved order parameters}

\subsection{Cahn-Hilliard revisited}

To illustrate our basic formalism, we take a look back at the foundations of phase field theories in the form of the equation of Cahn and Hilliard for the evolution of conserved order parameters. We consider a scalar order parameter $\rho$ (think of a density of particles) whose evolution fulfils a conservation law which we formulate in the 'hydrodynamic' form 
\begin{equation}
\partial_t \rho + \nabla \cdot(\rho \Bv) = 0,
\label{eq:cont1}
\end{equation}
where $\Bv$ is the particle transport velocity amd $\cdot$ denotes the inner product. The evolution of $\rho$ is supposed to minimize a free energy function $F$ which we take of the form 
\begin{equation}
F = \int_{\cal V} \left[\Phi(\rho) + \frac{\Gamma}{2} (\nabla\rho)^2\right] \diff^3 r.
\label{eq:F1}
\end{equation}
Here $\cal V$ denotes the domain of solution on which $\rho$ is defined, $\Psi$ is a free energy density and $\Gamma$ characterizes an interface energy penalty. We now evaluate the change of the free energy owing to the change of $\rho$, 
\begin{eqnarray}
d F/dt  &=& \int_{\cal V} \left[\frac{\partial\Psi}{\partial \rho} \frac{\partial\rho}{\partial t} + \Gamma \nabla \rho \cdot \nabla\frac{\partial\rho}{\partial t}\right] \diff^3 r
\nonumber\\
&=& \int_{\cal V} -\left[\frac{\partial\Phi}{\partial \rho} \nabla \cdot (\rho \Bv) + \Gamma \nabla \rho \cdot \nabla\cdot\nabla(\rho\Bv)\right] \diff^3 r.
\end{eqnarray}
where we have used Eq. (\ref{eq:cont1}). Now we carry out integrations by parts, integrating the first term on the right-hand-side once and the second term twice. Assuming bulk behavior where the fluxes $\rho\Bv$ and their derivatives can be set to zero on the surface of $\cal V$, we obtain 
\begin{equation}
d F/dt = \int_{\cal V} \rho \Bv \cdot \nabla  \left[\frac{\partial\Phi}{\partial \rho}  - \Gamma \Delta \rho \right] \diff^3 r.
\label{eq:dF2}
\end{equation}
Negative definiteness of the free energy derivative is ensured if we choose the transport velocity as
\begin{equation}
    \Bv = - M(\rho)\nabla \left[\frac{\partial\Psi}{\partial \rho}  - \Gamma \Delta \rho \right],
\end{equation}
where $M$ is a positively definite mobility function. Inserting into the continuity equation gives 
\begin{equation}
    \partial_t \rho = \nabla \cdot \left(\rho M(\rho)\nabla \left[\frac{\partial\Psi}{\partial \rho}  - \Gamma \Delta \rho \right]\right),
\end{equation}
which is the familiar Cahn-Hilliard equation. A similar procedures can be applied to any problem which describes the velocity-controlled transport of a conserved quantity, provided that the transport is driven by minimization of a free energy functional. We now apply this idea for the problem of dislocation motion. 

\subsection{Dislocation densities as conserved vectorial order parameters}

On the continuum level, dislocations are described by the Kr\"oner-Nye dislocation density tensor $\Balpha$ which defines the dislocation density as the curl of the plastic distortion, $\Balpha = - \nabla \times \Bbetapl$. For this quantity, Mura \cite{mura1963continuous} formulated the evolution equation 
\begin{equation}
\partial_t \Balpha = \nabla \times [\Bv \times \Balpha],
\label{eq:mura}
\end{equation}
where $\Bv$ is the dislocation velocity. The nature of this equation as a conservation law was already observed in the original paper of Mura. To relate it to a free energy functional, it is however useful to consider dislocation densities and dislocation velocities as slip system specific quantities, a consideration which is already implicit in the original work of Mura. In case of dislocations moving by slip on crystallographic slip systems, we may thus relate their motion to the respective thermodynamic driving forces, i.e., the resolved shear stresses on the various slip systems. 

To this end, we resolve $\Balpha$ into slip system specific contributions, $\Balpha = \sum_{\beta} \Balpha^{\beta}$, where $\beta$ is a slip system index and the $\Balpha^{\beta}$ can be written as tensor products of slip system specific dislocation density vectors and Burgers vectors:
\begin{equation} 
\Balpha^{\beta} = \Brho^{\beta} \otimes \Bb^{\beta}.
\end{equation}
Then, the dislocation density vectors obey the equations 
\begin{equation}
\partial_t \Brho^{\beta} = \nabla \times [\Bv^{\beta} \times \Brho^{\beta}].
\label{eq:rhovect}
\end{equation}
To proceed, we first focus on dislocation systems where, in each volume element, only dislocations of the same unit tangent vector $\Bt$ are present. In this case, the dislocation density vectors can be written as $\Brho = \rho \Bt$ where $\rho = |\Brho| = (\Brho.\Brho)^{1/2}$ has the meaning of a scalar dislocation length per unit volume. Such {\em unidirectional dislocation density fields} are considered in the works of El-Azab and co-workers \cite{xia2015computational,lin2020implementation} and in the field-based approach of Bertin \cite{bertin2019connecting} towards discrete dislocation simulation. 

\subsection{Unidirectional dislocation density fields}

To formulate a phase field equation for the evolution of unidirectional dislocation density fields, we first observe that for such fields, the slip system specific dislocation velocities $\Bv^{\beta}$ are, in case of dislocations moving by slip on slip planes with normal vectors $\Bn^{\beta}$, of the form
\begin{equation}
\Bv^{\beta} = v^{\beta} (\Bt^{\beta} \times \Bn^{\beta}),
\label{eq:rhovelo}
\end{equation}
This relationship allows us to re-write Eq. (\ref{eq:rhovect}) as
\begin{equation}
\partial_t \Brho^{\beta} = \nabla \times [  \rho^{\beta} v^{\beta} (\Bt^{\beta} \times \Bn^{\beta} \times \Bt^{\beta})] = - \Bn^{\beta} \times \nabla (\rho^{\beta} v^{\beta}).
\label{eq:rhovect1}
\end{equation}
where we have used that $\Bn^{\beta}$ and $\Bt^{\beta}$ are perpendicular vectors, hence $\Bt ^{\beta}\times \Bn^{\beta} \times \Bt^{\beta} = \Bn^{\beta}$, and that $\nabla\times \Bn^{\beta}=0$. For further use, we note that the scalar dislocation densities $\rho^{\beta} = |\Brho^{\beta}| = \Bt^{\beta}.\Brho^{\beta}$ fulfil the equation
\begin{equation}
\partial_t \rho^{\beta} = - \Bt^{\beta}.[\Bn^{\beta} \times \nabla (\rho^{\beta} v^{\beta})].
\label{eq:rhoscal0}
\end{equation}
or component-wise
\begin{eqnarray}
\partial_t \rho^{\beta} = - t^{\beta}_i \epsilon_{ijk} n^{\beta}_j \nabla_k (\rho^{\beta} v^{\beta})
= -  \epsilon_{ijk} n^{\beta}_j \nabla_k (t^{\beta}_i \rho^{\beta} v^{\beta}) +  \rho^{\beta} v^{\beta} \epsilon_{ijk} n^{\beta}_j \nabla_k t^{\beta}_i .
\label{eq:rhoscal1}
\end{eqnarray}
Thus, in vector form, 
\begin{equation}
\partial_t \rho^{\beta}  
= - [\Bn^{\beta} \times \nabla] \cdot (\Brho^{\beta} v^{\beta})] + q^{\beta} v^{\beta} ,
\label{eq:rhoscal2}
\end{equation}
where 
\begin{equation}
    q^{\beta} = k^{\beta} \rho^{\beta},\quad
    {\rm and} \quad k^{\beta} = (\Bn^{\beta} \times \nabla) \cdot \Bt^{\beta} =  \Bn^{\beta}\cdot[\nabla  \times \Bt^{\beta}]
\label{eq:kq}
\end{equation}
is the curl of the tangent vector in the slip plane, i.e., the local dislocation curvature. The variables $q^{\beta}$ have been denoted the curvature or loop densities. For later use we discuss the geometrical meaning of these quantities as established by Hochrainer \cite{hochrainer2014continuum}: The integral $n_{\cal V}^{\beta} = \int_{\cal V} q^{\beta} dV$ gives the number of oriented dislocation loops of slip system $\beta$ contained in the volume $\cal V$. In evaluating this number, the following conventions are used: 
\begin{itemize}
    \item A loop counts positive if the tangent vector $\Bt^{\beta}$ rotates counter-clockwise around $\Bn^{\beta}$.
    \item A positive loop expands under a positive shear stress.
    \item If a loop enters ${\cal V}$ at an angle $\phi_1$ and exits at angle $\phi_2$, then it makes a fractional contribution $(\phi_1-\phi_2)/(2\pi)$ to $n_{\cal V}^{\beta}$.  
\end{itemize}

From equation (\ref{eq:rhoscal2}) it is evident that the evolution equations of $\Brho^{\beta}$ for curved dislocations in general do not conserve the total line length per unit volume, and that the change in line length is associated with the presence of loops.   

Glide motion of dislocations is associated with the development of plastic eigenstrain, where the plastic strain rate is for glide motion of dislocations and single-valued dislocation fields given by
\begin{equation}
\dot{\Bepspl} = \sum_{\beta} \rho^{\beta}v^{\beta} \BM^{\beta}
\label{eq:Bepspl}
\end{equation}
with $\BM^{\beta}=(b^{\beta}\otimes n^{\beta}+
n^{\beta}\otimes b^{\beta})/2$. 

We now proceed to derive a phase-field type equation by formulating a free energy functional that accounts, on the one hand, for the dislocation related defect energy and, on the other hand, for the elastic energy associated with a generic plastic strain field caused by the motion of the dislocations. This functional yields the thermodynamic driving forces for the evolution of the dislocation system.  

We consider the free energy function in the form
\begin{equation}
F = \int_V (\Psi_{\rm D} + \Psi_{\rm E}) \diff V.
\end{equation}
The elastic energy density is taken in the standard elastic-plastic form
\begin{equation}
\Psi_{\rm 
E} = \frac{1}{2}(\Beps-\Bepspl):\CC:(\Beps-\Bepspl).\label{eq:PhiE}
\end{equation}
The defect energy density is assumed as the product of the total dislocation density $\rho = \sum_{\beta} \rho^{\beta}$ and 
the dislocation line energy,
\begin{equation}
\Psi_{\rm D} = A G b^2 \rho \ln (\rho/\rho_0) .
\end{equation}
Here, $A$ is a numerical factor of the order of one, $G$ the shear modulus of the material, and $b$ the length of the Burgers vectors $\Bb^{\beta}$. Associated with the defect energy density we define chemical potentials 
\begin{equation}
\mu_{\rho}^{\beta} = \partial \Psi_{\rm D}/\partial \rho^{\beta} = A G b^2 [\ln (\rho/\rho_0)+1]
\end{equation}
and stress-like variables ('back stress' $\tau_{\rm b}^{\beta}$ and 'curvature stress' $\tau_{\rm c}^{\beta}$) 
\begin{equation}
\tau_{\rm d}^{\beta} = - \frac{\Bt^{\beta}}{b} \cdot (\Bn^{\beta} \times \nabla) \mu_{\rho}^{\beta} = - \frac{A G b}{\rho} \Bt^{\beta} \cdot (\Bn^{\beta} \times \nabla) \rho\quad,\quad
\tau_{\rm c}^{\beta} = \frac{\mu_{\rho}^{\beta} k^{\beta}}{b}.
\label{eq:stressesuni}
\end{equation}
We proceed to formulate the evolution of the free energy function due to the motion of dislocations:
\begin{eqnarray}
\frac{{\diff} F}{{\diff} t} &=&  \int_V
\left[\sum_{\beta} \frac{\partial \Psi_{\rm D}}{\partial \rho^{\beta}} \partial_t \rho^{\beta} + \frac{\partial \Psi_{\rm E}}{\partial \Bepspl} \partial_t \Bepspl \right] \diff V\nonumber\\
&=& \sum_{\beta} \int_V \left[ - \mu^{\beta} \Bt^{\beta}\cdot(\Bn^{\beta} \times \nabla)(\rho^{\beta} v^{\beta})
-  \tau^{\beta} \rho^{\beta} b v^{\beta} \right]\diff V\nonumber\\
\end{eqnarray}
where the resolved shear stresses in the slip systems are defined as $\tau^{\beta} = (\Beps - \Bepspl):\CC:\BM^{\beta}/b$. We now perform a partial integration of the first term on the right-hand side. This gives
\begin{eqnarray}
\frac{{\diff} F}{{\diff} t} &=& \sum_{\beta}  \int_V
 \rho^{\beta} b v^{\beta} \left[ \tau^{\beta}_{\rm c} - \tau^{\beta}_{\rm b} - \tau^{\beta} \right]\diff V
\nonumber\\
&+& \sum_{\beta} \int_{\partial V}   \rho^{\beta} v^{\beta} \left[\Bt^{\beta} \cdot (\Bn^{\beta} \times \Bn^{\rm S})\right] \mu^{\beta} \diff A
\end{eqnarray}
where the second integral runs over the body surface and $\Bn^{\rm S}$ is the surface normal vector. 
To ensure negative definiteness of the free energy change, we require the velocites $v^{\beta}$ to be of the form
\begin{equation}
 v^{\beta} = \frac{B}{b} \left[\tau^{\beta}  + \tau^{\beta}_{\rm d} - \tau^{\beta}_{\rm c}\right]
\label{eq:TDcon}
\end{equation}
where $B$ is a positively definite mobility function. This form has the drawback that dislocations are moving even at arbitrarily low energetic driving forces, whereas in most realistic situations, this motion is hindered by friction-like terms due to Peierls stresses (lattice friction) or forest interactions. An alternative form which is equally consistent is given by 
\begin{equation}
 v^{\beta} = \frac{B}{b} \chi(\tau_*^{\beta},\tau_{\rm y}^{\beta}) \quad,\quad \tau_*^{\beta} = \tau^{\beta}  + \tau^{\beta}_{\rm d} - \tau^{\beta}_{\rm c}.
\label{eq:TDyield1}
\end{equation}
where $\tau_{\rm y}^{\beta}$ is a yield stress (for forest hardening, $\tau_{\rm y}^{\beta} = \alpha G b \sqrt{\rho}$) and the mobility function is 
\begin{eqnarray}
\chi(\tau_*^{\beta},\tau_{\rm y}^{\beta}) &=& \tau_*^{\beta} - \tau_{\rm y} {\rm sign}(\tau_*^{\beta})\quad,\quad
|\tau_*^{\beta}| - \tau_{\rm y} > 0,\nonumber\\
\chi(\tau_*^{\beta},\tau_{\rm y}^{\beta}) &=& 0 \quad,\quad 
|\tau_*^{\beta}| - \tau_{\rm y} \le 0.
\label{eq:TDyield2}
\end{eqnarray}

Satisfying the condition for negative definiteness of the bulk energy change does in general not set the surface integral to negative values. Therefore, in addition we must impose at the surface either the geometrical boundary condition 
\begin{equation}
\Bt^{\beta} \cdot  (\Bn^{\beta} \times \Bn^{\rm S})= 0 \quad\forall \Br \in \partial V.
\label{eq:surface}
\end{equation}
or the no-flux boundary condition $\rho^{\beta}v^{\beta} = 0$. 
In simple words, Eq. (\ref{eq:surface}) means that at the surface the dislocation tangent vector must be perpendicular to the line of intersection between the surface and the slip plane. This imposes in fact conditions on the plastic slip gradient - the surface normal slip gradients must be zero. 

Both types of boundary conditions have been considered in the past when modelling size dependent mechanical behavior using dislocation density dynamics. For example, 'no flux' boundary conditions were considered by Wu et al \cite{wu2017continuum} at the interface between an fcc matrix and plastically non-deformable inclusions to understand size dependent plasticity of superalloys, and by \cite{luo2023computationally} at grain boundaries to analyze grain size effects in Mg. 'No slip gradient' boundary conditions were considered by Zaiser et. al. \cite{zaiser2007modelling,sandfeld2010numerical} to explain size dependent behavior in microbending of thin films. 

Thus, our thermodynamic considerations not only allow to establish the form of the dislocation velocities, but also yield boundary conditions for slip rates and/or slip gradients on the sample surface.

\subsection{Multidirectional dislocation density fields}

To fully exploit the power of a continuum approach, it is desirable to extend the approach towards scales where an elementary volume contains dislocation of more than one orientation on a given slip system. This is, in particular, essential if we want to model dislocation behavior on the grain and multi-grain scales. To deal with such situations, Hochrainer and co-workers \cite{2007_PM_Hochrainera} proposed a theory based on considering, in each elementary volume, continuous orientation-specific density functions $\rho^{\beta}(\phi)$ where we may understand $\phi$ as the angle between dislocation line direction and Burgers vector. The structure of this higher-dimensional theory closely resembles Eq. (\ref{eq:rhovect}), with a velocity that is generalized to include transport in orientation space (i.e. dislocation rotation). The theory was applied to a range of problems in size dependent plasticity \cite{zaiser2007modelling,sandfeld2010numerical}, but it suffers from a high computational cost which has led to a search for simplified formulations. In the following we outline such a formulation which again considers the scalar dislocation densities $\rho^{\beta}$ and density vectors $\Brho^{\beta}$, but envisages elementary volumes containing dislocations of multiple orientations. In this case, the densities fulfil the triangular inequality $\Brho^{\beta}.\Brho^{\beta} \le (\rho^{\beta})^2$. 

The dislocation arrangement in an elementary volume is envisaged as the superposition of unidirectional dislocation density fields $\rho^{\beta}_{\Bt}, \Brho^{\beta}_{\Bt}$ which we now label with the subscript $\Bt$ to indicate that they refer to a single tangent vector. We introduce normalized orientation distribution functions $p^{\beta}(\Bt^{\beta})$ where $\Bt^{\beta}(\omega)$ are vectors on the unit circle in the slip plane of slip system $\beta$ (this is equivalent to considering a distribution of orientation angles $\omega$ but renders the presentation more concise). The unidirectional density functions for orientation $\Bt^{\beta}$ fulfill the relationship $\rho^{\beta}_{\Bt} = \rho^{\beta} p^{\beta}(\Bt^{\beta})$ where $ \rho^{\beta}$ is the multidirectional density. The multidirectional dislocation densities and dislocation density vectors then fulfill the relations
\begin{eqnarray}
\Brho^{\beta}(\Br) = \int \Brho^{\beta}_{\Bt} \diff \omega  = \int \rho^{\beta}_{\Bt} \Bt^{\beta}\diff \omega = \rho^{\beta} \int p^{\beta}(\Bt^{\beta}) \Bt^{\beta} \diff \omega
\label{eq:Brhobar}
\end{eqnarray}
In addition, we define GND densities $\kappa^{\beta}$ and GND tangent vectors $\Bt^{\beta}_{\kappa}$ via
\begin{equation}
    \kappa^{\beta} = ||\Brho^{\beta}||\quad,\quad
    \Bt^{\beta}_{\kappa} = \frac{\Brho^{\beta}}{\kappa^{\beta}}. 
\end{equation}
Note that the GND density is here a positively definite quantity, different from 2D models of dislocation systems where $\kappa$ carries a sign. In the present 3D formalism, the sign of the GNDs is implicitly defined by the direction of their tangent vector. 

Using the above notations, we use the principle of maximum entropy in order to estimate the orientation distribution functions in terms of $\rho^{\beta}$, $\kappa^{\beta}$ and $\Bt^{\beta}_{\kappa}$. Using results of Monavari et. al. \cite{monavari2014comparison,monavari2016continuum}, the maximum entropy estimates of the orientation distributions $p(\Bt^{\beta})$ are in our notation given by
\begin{equation}
p(\Bt^{\beta}) =  {\cal N}(\Lambda^{\beta}) \exp\left[-\Lambda^{\beta}(\Bt^{\beta}\cdot \Bt_{\kappa}^{\beta})\right]\;,\;
{\cal N} = 2 \pi I_0(\Lambda^{\beta})\;,\;
\frac{\kappa^{\beta}}{\rho^{\beta}}=\frac{ I_1(\Lambda^{\beta})}{ I_0(\Lambda^{\beta})},
\label{eq:maxent}
\end{equation}
where the $I_n$ are modified Bessel functions. We now average the evolution equations of single-orientation dislocation fields, Eqs. (\ref{eq:rhovect1},\ref{eq:rhoscal2}), over these orientation distributions.  

\subsubsection{Kinematic equations for multidirectional density functions}

When averaging the kinematic evolution equations of unidirectional dislocation density fields, we account for the fact that dislocations of different orientation may have different velocities, even when they are located in the same elementary volume and subject to the same local shear stress. This is true even if the dislocation mobility is independent of dislocation character (edge vs screw). To understand the reason, consider a volume containing straight parallel dislocations of tangent vectors $\Bt^{\pm} = \pm \Be_x$. Assume there are more dislocations of 'positive' orientation $\Bt^+$ than of 'negative' orientation $\Bt^-$. Two dislocations of tangents $\Bt^+$ and $\Bt^-$ may form a temporarily immobile dipole. It is then clear that all negative dislocations can be trapped into dipoles but a part of the positive ones will remain mobile, leading to a higher mean velocity for the positive dislocations. This observation has long been taken into account in continuum descriptions of two-dimensional dislocation systems consisting of straight parallel dislocations \cite{2016_PRB_Valdenaire, groma2016dislocation,wu2018instability,wu2021cell}. 
In our three-dimensional formulation, the distinction of 'positive' and 'negative' dislocations is replaced by a continuous orientation dependence of the dislocation velocity. Specifically, we assume the velocity function to depend on the tangent vector via
\begin{equation}
    v(\Bt) = v_0 + v_{1} (\Bt\cdot \Bt_{\kappa})
    \label{eq:v12}
\end{equation}
This form is equivalent to the one assumed by Groma and co-workers \cite{groma2021dynamics} and corresponds to a first-order Fourier expansion of a generic dislocation velocity spectrum. For a system of straight parallel dislocations, it is equivalent to the assumption of two different velocities $v^{\pm} = v_0 \pm v_{1}$ depending on whether a dislocation has the majority or minority orientation. 

We now first evaluate the plastic strain rates, by considering the expression for single-valued dislocation densities and integrating this over all directions of the tangent vector. We obtain
\begin{eqnarray}
\dot{\gamma}^{\beta} &=& b \int \rho_{\BT}^{\beta} v(\Bt^{\beta}) \diff \omega
\nonumber\\
&=& b (\rho^{\beta} v_0^{\beta} + \Brho^{\beta}\cdot \Bv^{\beta}_1),
\end{eqnarray}
where we have introduced the notation $\Bv^{\beta}_1 = \Bt^{\beta}_{\kappa} v^{\beta}_{1}$. Accordingly, the evolution equation of the GND density vector follows as 
\begin{equation}
\partial_t \Brho^{\beta} = - [\Bn^{\beta} \times \nabla] \cdot [\rho^{\beta} v_0^{\beta} + \Brho^{\beta}\cdot \Bv^{\beta}_1]
\end{equation}
Next, we look at the directional average of the evolution equation for the scalar dislocation density, Eq. (\ref{eq:rhoscal2}). The flux term in this equation can after angular integration be re-written as follows:
\begin{eqnarray}
&-& \int [\Bn^{\beta} \times \nabla] \cdot [\rho^{\beta}_{\Bt} \Bt^{\beta} v(\Bt^{\beta}) ] \diff \omega\nonumber\\
&=& -[\Bn^{\beta} \times \nabla] \cdot \int \rho^{\beta}_{\Bt}\Bt^{\beta} [v^{\beta}_0 
+ \Bt^{\beta}\cdot \Bv^{\beta}_{1}]\diff \omega \nonumber\\
&=& -[\Bn^{\beta} \times \nabla] \cdot \left[ \Brho^{\beta} v^{\beta}_0 
+ \Brho^{(2),\beta}\cdot \Bv^{\beta}_1 \right]
\end{eqnarray}
Here, a new field variable emerges which is the second-order alignment tensor of the orientation distribution \cite{hochrainer2015multipole}. We evaluate this tensor using Eq. (\ref{eq:maxent}): 
\begin{eqnarray}
\Brho^{(2),\beta} &=& \int \rho^{\beta}_{\Bt}[\Bt^{\beta}  \otimes \Bt^{\beta}] \diff \omega =
\rho^{\beta}\int p(\Bt^{\beta})[\Bt^{\beta}  \otimes \Bt^{\beta}] \diff \omega  \nonumber\\
&\approx&  \frac{\rho^{\beta}}{2}\left\{(1+\Phi^{\beta}) [\Bt^{\beta}_{\kappa}  \otimes \Bt^{\beta}_{\kappa}]+(1-\Phi^{\beta}) [\Bg^{\beta}_{\kappa}  \otimes \Bg^{\beta}_{\kappa}]\right\}.
\label{eq:rhotwo}
\end{eqnarray}
Here, $\Bg^{\beta}_{\kappa} = \Bn^{\beta}\times \Bt^{\beta}_{\kappa} $ is the unit vector in GND glide direction. The function $\Phi^{\beta}$ depends on the ratio $\kappa^{\beta}/\rho^{\beta}$. A polynomial approximation is given by 
\begin{equation}
\Phi^{\beta}\left( \frac{\kappa^{\beta}}{\rho^{\beta}}\right) \approx \frac{1}{2}\left[\left( \frac{\kappa^{\beta}}{\rho^{\beta}}\right)^2+\left( \frac{\kappa^{\beta}}{\rho^{\beta}}\right)^6\right]
\end{equation}
The next step is to average the term involving dislocation curvature:
\begin{equation}
\int \rho^{\beta}_{\Bt} [(\Bn^{\beta} \times \nabla)\Bt^{\beta}] v(\Bt^{\beta}) \diff \omega
= \rho^{\beta}\int p(\Bt^{\beta}) [(\Bn^{\beta} \times \nabla)\Bt^{\beta}] (v^{\beta}_0 + \Bt^{\beta}\cdot \Bv^{\beta}_1) \diff \omega 
\end{equation}
We first consider the term involving $\Bv^{\beta}$. This term can be related to the second-order alignment tensor $\Brho^{2}$ in terms of the so-called curvature vector introduced by Hochrainer \cite{hochrainer2015multipole}:
\begin{equation}
\Bq^{\beta} = \nabla \cdot \Brho^{(2),\beta}.
\end{equation}
Using the relation $\nabla \cdot (\rho^{\beta}_{\Bt} \Bt^{\beta}) = 0$ (solenoidality of the dislocation density vectors), we can show that 
\begin{equation}
\int \rho_{\Bt} [(\Bn^{\beta} \times \nabla)\Bt^{\beta}]   \Bt^{\beta}\cdot \Bv^{\beta}_1 \diff \omega = - (\Bn^{\beta} \times \Bq^{\beta})\cdot \Bv^{\beta}_1 = - (\Bn^{\beta} \times \nabla \Brho^{(2),\beta}) \cdot \Bv^{\beta}_1.
\end{equation}
Thus, this average can be expressed in terms of the dislocation density variables. More problematic is the second term
\begin{equation}
\int \rho^{\beta}_{\Bt} [(\Bn^{\beta} \times \nabla)\Bt^{\beta}] v^{\beta}_0 \diff \omega  =: q^{\beta}_0 v^{\beta}_0,
\end{equation}
where we have introduced the mean curvature  
\begin{equation}
q^{\beta}_0 = \int \rho^{\beta}_{\Bt} [(\Bn^{\beta} \times \nabla)\Bt^{\beta}]\diff \omega 
\end{equation}
As observed by Hochrainer \cite{hochrainer2015multipole}, this quantity can be interpreted as a volume density of dislocation loops. Assembling all terms, the evolution equation for the scalar density fields $\rho^{\beta}$ can after some re-writing be cast into the form
\begin{eqnarray}
\partial_t \rho^{\beta} &=& - [\Bn^{\beta} \times \nabla] \cdot [\Brho^{\beta} v^{\beta}_0 + \rho^{\beta}\cdot \Bv^{\beta}_1]\nonumber\\
&+& q^{\beta}_0 v^{\beta}_0 + \frac{1-\Phi^{\beta}}{2}\rho^{\beta}[\Bn^{\beta} \times \nabla] \Bv^{\beta}_1 +
\Phi^{\beta}\rho^{\beta} k^{\beta}_{\kappa} \Bt^{\beta}_{\kappa} \cdot \Bv^{\beta}_1,
\label{eq:rhomulti}
\end{eqnarray}
where we have introduced the GND curvature 
\begin{equation}
k_{\kappa}^{\beta} = [\Bn^{\beta} \times \nabla]\Bt_{\kappa}^{\beta}.
\end{equation}
We note that an equation analogous to Eq. (\ref{eq:rhomulti}) was derived by Hochrainer and collaborators 
\cite{groma2021dynamics}. In that work the derivation was restricted to weakly anisotropic dislocation arrangements ($\Phi \ll 1$) and accordingly the GND curvature related term is missing. 

In the works of Hochrainer and collaborators (see e.g. \cite{groma2021dynamics}), the loop densities $q^{\beta}_0$ are treated as independent density-like variables obeying their own transport equations. 
Here we choose a different path and specify the $q^{\beta}$ constitutively, using the following principles: (i) In the limit of single-valued dislocation density fields we must recover the curvature of the GND tangent vector as given by Eq. (\ref{eq:kq}); (ii) in the general case, the expression used for GND curvature must be consistent with generic scaling properties of dislocation fields \cite{zaiser2014scaling} 
We assume the average curvature density as the weighted average of GND and SSD contributions, using $\Phi$ as the weighting function:
\begin{equation}
    q^{\beta}_0 = \rho^{\beta} [\Phi k_{\kappa}^{\beta}  + (1-\Phi) k^{\beta}_{\rm SSD}]
    \quad,\quad k^{\beta}_{\rm SSD} = \xi \;\rm{sign}(\tau^{\beta})\sqrt{\rho}
    \label{eq:kbeta}
\end{equation}
This implies that the curvature radii of SSD loops are assumed proportional to the mean dislocation spacing, and the loops are oriented such that they expand under the shear stress acting on the respective slip system. 

\subsection{Driving forces and velocities for multidirectional density functions}

The derivation of velocities from the free energy functional proceeds as for the case of single-valued dislocation density fields. Since we now use two fields to characterize the dislocation microstructure, we choose the defect related part of the free energy density as \cite{zaiser2015local,2016_JMPS_Hochrainer}
\begin{equation}
\Psi_{\rm D} = \sum_{\beta}\left(A G b^2 \rho^{\beta} \ln (\rho/\rho_0) + D G b^2 \frac{\Brho^{\beta}\cdot \Brho^{\beta}}{2 \rho}\right).
\label{eq:energycdd2}
\end{equation}
Note that we do not include a quadratic dependency of the defect energy on line curvature as proposed by Hochrainer \cite{2016_JMPS_Hochrainer}, for which we see no clear physical motivation. Associated with the free energy density, we define quasi-chemical potentials for the dislocation density variables: 
\begin{eqnarray}
\mu^{\beta}_{\rho} = \frac{\partial \Psi_{\rm D}}{\partial \rho^{\beta}} = G b^2 \left[A[\ln(\rho/\rho_0)+1] - B \frac{\Brho^{\beta}\cdot \Brho^{\beta}}{2\rho^2}\right],\nonumber\\
\Bmu^{\beta}_{\kappa} = \frac{\partial \Psi_{\rm D}}{\partial \Brho^{\beta}} = D G b^2  \frac{\Brho^{\beta}}{\rho},
\end{eqnarray}  
where $\Bmu^{\beta}_{\kappa}$ is a vector compiling the potentials $(\mu^{\beta}_{\kappa})_i = \partial \Phi_{\rm D}/\partial \Brho^{\beta}_i$. Moreover, we define shear stress-like quantities as follows
\begin{eqnarray}
\tau^{\beta}_{\rm b} &= - (\Bn^{\beta} \times \nabla) \Bmu^{\beta}_{\kappa}
\;,\; 
\Btau^{\beta}_{\rm d,0} = - (\Bn^{\beta} \times \nabla) \mu^{\beta}_{\rho},\nonumber\\
\Btau^{\beta}_{\rm d,1} &= \Btau^{\beta}_{\rm d,0} + \frac{1}{2\rho^{\beta}} (\Bn^{\beta} \times \nabla)[ (1-\Phi^{\beta})\rho^{\beta}\mu^{\beta}_{\rho}], \nonumber\\
\tau^{\beta}_{\rm c} &= \frac{\mu_{\rho}^{\beta} q^{\beta}_0}{\rho^{\beta}}\;,\;
\Btau^{\beta}_{\rm c} = \Phi^{\beta}\rho^{\beta} k^{\beta}_{\kappa} \Bt^{\beta}_{\kappa}.
\end{eqnarray} 
With these notations, the evolution of the free energy function is written as
\begin{eqnarray}
\frac{{\diff} F}{{\diff} t} =  \sum_{\beta} \int_V & - \mu_{\kappa}^{\beta} [\Bn^{\beta} \times \nabla] \cdot [\rho^{\beta} v_0^{\beta} + \Brho^{\beta}\cdot \Bv^{\beta}_1] \nonumber\\
& - \Bmu_{\rho}^{\beta} [\Bn^{\beta} \times \nabla] \cdot [\Brho^{\beta} v^{\beta}_0 + \rho^{\beta}\cdot \Bv^{\beta}_1]\nonumber\\
& + \Bmu_{\rho}^{\beta} [q^{\beta}_0 v^{\beta}_0 + \frac{1-\Phi^{\beta}}{2}\rho^{\beta}[\Bn^{\beta} \times \nabla] \Bv^{\beta}_1 + \Phi^{\beta}\rho^{\beta} k^{\beta}_{\kappa} \Bt^{\beta}_{\kappa} \cdot \Bv^{\beta}_1]\nonumber\\
& - \tau^{\beta} b [\rho^{\beta} v_0^{\beta} + \Brho^{\beta}\cdot \Bv^{\beta}_1]
\diff V.
\end{eqnarray}
Upon partial integration one gets
\begin{eqnarray}
\frac{{\diff} F}{{\diff} t} =  \sum_{\beta} \int_V & \left\{- (\tau^{\beta} +\tau^{\beta}_{\rm b})  
[\rho^{\beta} v^{\beta}_0 + \Brho^{\beta}\cdot \Bv^{\beta}_1] - \Btau_{\rm d,0}^{\beta} 
[\Brho^{\beta} v^{\beta}_0 + \rho^{\beta} \Bv^{\beta}_1] \right. \nonumber\\
& \left. + \rho^{\beta} \tau^{\beta}_{\rm c} v^{\beta}_0 b + \rho^{\beta} (\Btau_{\rm d,1}-\Btau_{\rm d,0}) 
\cdot \Bv^{\beta}_1 b + \rho^{\beta} \Btau^{\beta}_{\rm c} \cdot \Bv^{\beta}_1 b \right\}\diff V.
\end{eqnarray}
Thermodynamic consistency requires the term in brackets to be negatively definite. Sorting the terms one finds that this condition reads 
\begin{eqnarray}
&\left(\tau^{\beta} +\tau^{\beta}_{\rm b} + \frac{\Brho^{\beta}}{\rho^{\beta}}\cdot \Btau_{\rm d,0}^{\beta} -
\tau^{\beta}_{\rm c} \right)v^{\beta}_0 \nonumber\\
+ &\left(\frac{\Brho^{\beta}}{\rho^{\beta}}(\tau^{\beta} +\tau^{\beta}_{\rm b}) -\Btau^{\beta}_{\rm d,1} - \Btau_{\rm c}^{\beta} \right) \cdot \Bv^{\beta}_1 \ge 0
\label{eq:TDconmult1}
\end{eqnarray}
We shall now consider the structure of dislocation velocity functions in two limit cases, namely (1) the limit of unidirectional dislocation systems, which must be recovered from the more general model if $\Brho^{\beta} = \rho^{\beta} \Bt^{\beta}_{\kappa}$, and (2) the limit of weakly anisotropic dislocation systems where $\Brho^{\beta} \ll\rho^{\beta} \Bt^{\beta}_{\kappa}$ treated in Ref. \cite{groma2021dynamics}.

\subsubsection{Unidirectional limit}

In the uni-directional limit $\Brho^{\beta} = \rho^{\beta} \Bt^{\beta}_{\kappa}$, $\Phi = 1$ and hence the stress contribution $\Btau^{\beta}_{\rm d,0} = \Btau^{\beta}_{\rm d,1}$ are identical. Moreover, the curvature related stresses become equivalent, $\Btau_{\rm c}^{\beta} = \Bt^{\beta}_{\kappa} \tau_{\rm c}^{\beta}$. Hence, the second term on the rhs. of Eq. (\ref{eq:TDcon}) becomes equivalent to the first term multiplied with $\Bt^{\beta}_{\kappa}$ and we can add up both terms to obtain  
\begin{eqnarray}
&\left(\tau^{\beta} +\tau^{\beta}_{\rm b} + \Bt^{\beta}_{\kappa}\cdot \Btau_{\rm d}^{\beta} -
\tau^{\beta}_{\rm c} \right)(v^{\beta}_0 + v^{\beta}_{1}) \ge 0
\label{eq:TDconmult2}
\end{eqnarray}
We now note that, if $\Brho^{\beta} = \rho^{\beta} \Bt^{\beta}_{\kappa}$, then $\Bt^{\beta}_{\kappa}\cdot \Btau_{\rm d}^{\beta} = \tau_{\rm d}^{\beta}$ as given by Eq. (\ref{eq:stressesuni}). Moreover, in the same limit, $\tau_{\rm b}^{\beta} = A \mu b k_{\kappa}^{\beta}$ can be absorbed into the curvature stress. Hence, Eq. (\ref{eq:TDconmult2}) becomes identical to Eq. (\ref{eq:TDcon}) for the unidirectional case. Accordingly, the flow rule for calculating the velocities can be taken in the form given by Eqs. (\ref{eq:TDyield1},\ref{eq:TDyield2}).

\subsubsection{Weakly anisotropic case}

The formulation of velocity laws for the weakly anisotropic case was considered in Ref. \cite{groma2021dynamics}, where the authors adjust their velocity laws such as to match the form derived for 2D dislocation systems using direct averaging of the Peach-Koehler forces \cite{groma2016dislocation}. In our terminology, the resulting relations are given by
\begin{equation}
v_0^{\beta} = \frac{B}{b}\chi(\tau^{\beta}_*,\tau^{\beta}_{\rm y})\quad,\quad
v_1^{\beta} = \frac{B}{b}\Bt^{\beta}_{\kappa}\cdot \left(\frac{\Brho^{\beta}}{\rho^{\beta}}[1-\chi(\tau^{\beta}_*,\tau^{\beta}_{\rm y})] + \Btau^{\beta}_{\rm d,1}\right),
\label{eq:TDyield2}
\end{equation}
where $\tau^* = \tau^{\beta} + \tau^{\beta}_{\rm b}$ and the yield stress must be chosen such that in the flowing phase the driving force $\tau^*$ is always high enough to ensure that the energy required for dislocation loop expansion \cite{wu2022thermodynamic}. The ensuing relations as well as the proof of consistency of Eq. (\ref{eq:TDyield2}) with Eq. (\ref{eq:TDconmult1}) are somewhat lengthy, we therefore refer the reader to Ref. \cite{groma2021dynamics}.

\subsection{Evaluation of long-range dislocation interactions in bulk systems}

Irrespective whether we are dealing with single-valued or multiple-valued dislocation density fields, solution of the equations of evolution includes the solution of an elastic-plastic problem, based on the balance equations of elasto-plasticity in conjunction with the free energy density, Eq. (\ref{eq:PhiE}) where the plastic strain derives from Eq. (\ref{eq:Bepspl}). For finite bodies, this can be done using standard methodology of crystal plasticity. In infinite bodies, an alternative approach is available based on the observation that internal stresses in an infinite body can be expressed in terms of  line integrals over closed dislocation loops. In the present continuum formulation, all GNDs (and in case of single valued dislocations, all dislocations) fall into this category, which allows us to express the internal stress field as an integral over the field $\Brho$. Specifically, 
we envisage the stress $\Bsigma = \Bsigma^{\rm ext} + \Bsigma^{\rm int}$ in an infinite body as the sum of a constant external stress $\Bsigma^{\rm ext}$, which arises from tractions applied on the infinite contour, and a space dependent internal stress field $\Bsigma^{\rm int}(\Br)$ which can be evaluated from the dislocation density vector field $\Brho$. Using results given by Cai et. al. \cite{cai2006non} we write the internal stress tensor as
\begin{equation}
\Bsigma^{\rm int}(\Br) = \sum_{\beta}
\int_V \Bb^{\beta}\cdot\BG(\Br-\Br')\cdot \Brho^{\beta}(\Br')\diff^3 r'.
\end{equation}
or in coordinate notation 
\begin{equation}
\sigma_{kl}^{\rm int} = \sum_{\beta}
\int_V b^{\beta}_o G_{oklm}(\Br-\Br')\cdot \rho^{\beta}_m(\Br')\diff^3 r'.
\end{equation}
where the tensor components $G_{oklm}$ are given by
\begin{eqnarray}
\BG_{oklm}(\vec{r})
&=& -\frac{\mu}{8\pi} \left\{  \frac{2}{1-\nu}\left( \frac{\partial^3 R}{\partial r_n \partial r_k \partial r_l} - \delta_{kl} \frac{\partial}{\partial r_n}\nabla^2R \right) \epsilon_{nom} \right. \nonumber\\
&+& \left.\left(\frac{\partial}{\partial r_n} \nabla^2 R\right) \left[\epsilon_{nok}\, \delta_{lm}    + \epsilon_{nol}\,\delta_{km} \right] \right\}, 
\label{eq:3Dstress}
\end{eqnarray}
where $R=|\Br - \Br'|$. 

For later use, we note that the elastic kernel allows for two types of {\em soft modes}, i.e., heterogeneous GND (or equivalently, strain) patterns that are not associated with any long-range internal stresses: 
\begin{itemize} 
\item{\em Tilt walls} are associated with wave-vectors that are parallel to the Burgers vector but perpendicular to the line direction of the GNDs, in other words, they can be thought of as composed of walls of vertically stacked edge dislocations. In this case, in Eq. (\ref{eq:3Dstress}), $k_n b_o = bk \delta_{no}$ and hence $\sigma_{kl} \propto \epsilon_{nok} \delta_{no} = 0$.
\item{\em Twist walls} are associated with wave-vectors that are perpendicular to the Burgers vectors of two mutually perpendicular sets of screw-type GNDs, in other words, they can be thought of as composed of walls of perpendicular screw dislocations. In this case, in Eq. (\ref{eq:3Dstress}), $b_o\rho_m^{\beta} = b\rho^{\beta} \delta_{no}$ and therefore the contribution of the first ('edge') term in the integral vanishes. Furthermore, for two sets of perpendicular screw dislocations where $\kappa^{(1)} = \kappa^{(2)} = \kappa$, $b_o \rho_k^{(1)} = b\kappa \delta_{ok}$ and $b_o \rho_l^{(1)} = b\kappa \delta_{ol}$, hence 
$\sigma_{kl} \propto (\epsilon_{nok} \delta_{ol} + \epsilon_{nol} \delta{ok}) = \epsilon_{nkl}+\epsilon_{nlk} = 0$. 
\end{itemize}

\section{Application to dislocation patterning in crystals with B1 lattice structure}

For illustrating the application of our phase field approach to the problem of dislocation patterning, we consider the formation of dislocation patterns in uni-axial and cyclic deformation of single crystals with B1(NaCl) lattice structure which are deforming in tension in a high-symmetry [100] orientation. Dislocation patterns in such crystals show many analogies with those in fcc crystals deforming in the same orientation. This includes the formation of cell structures in uni-axial deformation \cite{wu2021cell} and of labyrinth structures in cyclic deformation. We consider an infinite crystal with B1 lattice structure that is subject to a uni-axial stress $\Bsigma^{\rm ext} = \sigma \Be_x \otimes \Be_x$. Thus, there are four active and two inactive slip systems. The slip vectors $\Bs^i = \Bb^i/b$ and slip plane normal vectors $\Bn^i$ of the active slip systems are given by 
\begin{eqnarray}
        \Bs^1 = \frac{1}{\sqrt{2}}[110]\quad,\quad \Bn^1 = \frac{1}{\sqrt{2}}[1-10]\nonumber\\
        \Bs^2 = \frac{1}{\sqrt{2}}[1-10]\quad,\quad \Bn^2= \frac{1}{\sqrt{2}}[110]\nonumber\\    
        \Bs^3 = \frac{1}{\sqrt{2}}[101]\quad,\quad \Bn^3 = \frac{1}{\sqrt{2}}[10-1]\nonumber\\
        \Bs^4 = \frac{1}{\sqrt{2}}[10-1]\quad,\quad \Bn^4 = \frac{1}{\sqrt{2}}[101]\nonumber\\
\end{eqnarray}
The resolved shear stress on the active slip systems is $\tau^{\rm ext} = \sigma/M$ where $M = 2$. In the following we shall assume, until otherwise stated, a homogeneous initial dislocation density of $\rho_(0) = 10^{12}$ m$^{-2}$ on each of the active slip systems. The deformation state is fully homogeneous, i.e., the GND vector densities are identically zero, $\Brho^i = 0 \forall i$. We shall use the non-dimensional parameters $\alpha = 0.3, D=0.2, A=0.2$ as also considered in previous work \cite{wu2021cell}. The curvature parameter $\xi$ is set to $\xi = 0.03$ which implies that we assume an initial SSD loop radius of $30 \mu$m. Material parameters are for NaCl: $C_11 = , C_21=C_44=, b=2.5 \AA$.  

\begin{figure}[htpb]  \centering
    \includegraphics[width = \textwidth]{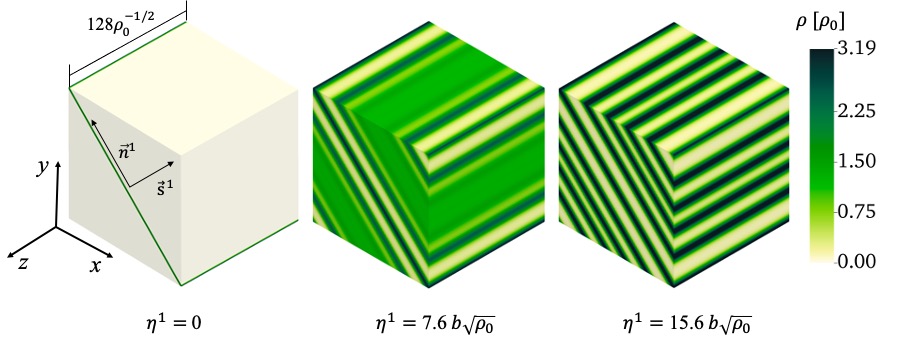}
    \includegraphics[width = \textwidth]{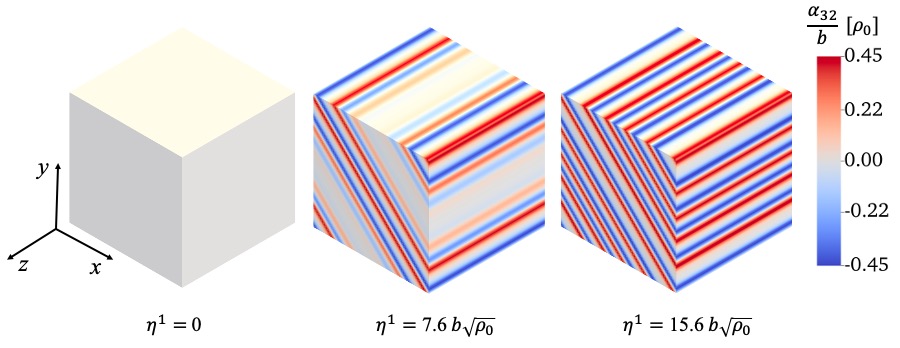}
    \caption{Formation of wall patterns in single slip; top: total dislocation density at different shear strains, bottom: dislocation density tensor component $\alpha_{32}=\alpha_{31}$, all other components are zero.}
    \label{fig:walls1s}
\end{figure}

Patterns emerge when inhomogeneous perturbations are present such that $\rho_1 = \rho_0 + \delta \rho_1(\Br)$. We consider two types of perturbations of the dislocation densities  $\rho^i$: First, we envisage planar density perturbations that are localized within a distance of one average dislocation spacing around the plane $\Bn_{\delta}\Br = 0$:
\begin{equation}
    \delta \rho^i(\Br) = \rho_0 \epsilon \exp\left(-\Br\Bn_{\delta} \rho_0\right))
\end{equation}
where the orientation of the perturbation imposes constraints on the ensuing patterns. The perturbation magnitude is chosen as $\epsilon = 0.01$. We use this type of initial perturbation to study the formation of simple wall and 2D cell structures as demonstrated in previous 2D models \cite{groma2016dislocation,wu2018instability,wu2021cell}. To this end we first consider situations where one slip system is active which we equate without loss of generality with slip system 1. We consider an initially localized perturbation of the density of the active slip system,  where the perturbation is localized around the plane $\Bn_{\delta} = \Bb^1/b$, i.e., it corresponds to an embryonic wall perpendicular to the slip direction, i.e., dislocations aligned with this wall in the $\Bn^1$ slip plane have edge orientation. In this case, the system develops a pattern of evenly spaced parallel walls that 'emanate' from the initial perturbation, as previously reported in 2D simulations of single slip deformation \cite{wu2018instability}, see \figref{fig:walls1s}. This figure shows the patterning of the total dislocation density alongside with the components of the dislocation density tensor 
\begin{equation}
    \Balpha = \sum_{\beta} \Brho^{\beta} \otimes \Bb^{\beta}
\end{equation}
which we divide by $b$ to facilitate comparison with the dislocation densities; thus, in \figref{fig:walls1s} where $\Brho = \kappa \Be_z$, the non-zero components fulfill the relationship $\alpha_{32} = \alpha_{31} = \kappa/\sqrt{2}$. The process that leads to formation of a fully developed pattern requires a strain of about 15 $b\sqrt{\rho_0}$ (note that $b\sqrt{\rho}$ is the strain accomplished when each dislocation moves on average one dislocation spacing), which is to be expected since the pattern wavelength is of the order of 25 dislocation spacings.

If the plane of the initial perturbation is parallel to the slip plane, $\Bn_{\delta} = \Bn^1$, no patterns can form. The same is true if the initial perturbation is in the plane $\Bn_{\delta} = \Bb^1\times \Bn^1/b$ such that dislocations aligned with the embryonic wall have screw orientation, as screw dislocation walls would create large stress fields. 

\begin{figure}[htpb]  \centering
    \includegraphics[width = \textwidth]{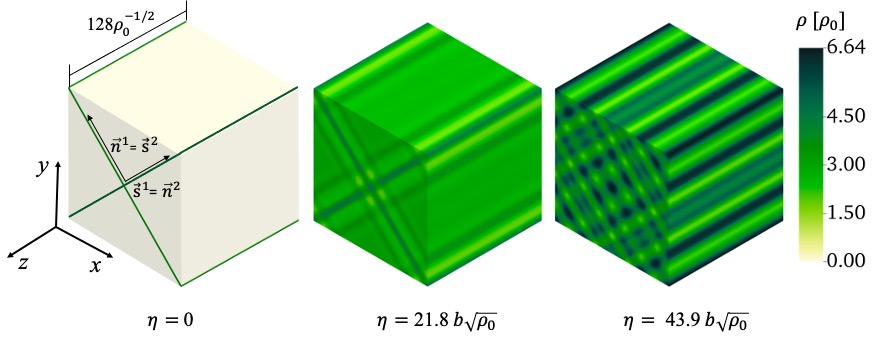}
    \includegraphics[width = \textwidth]{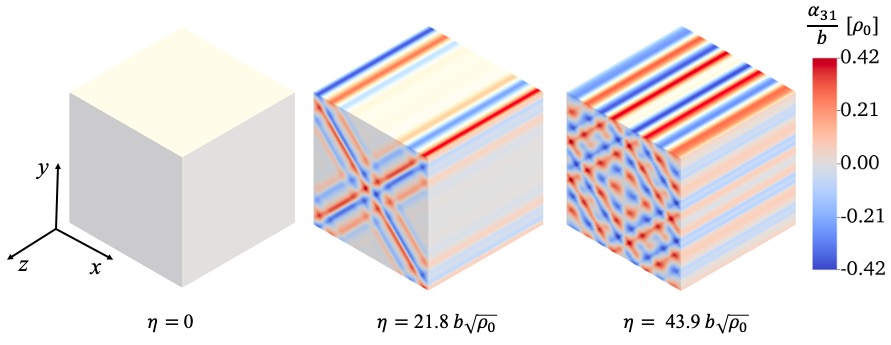}
    \caption{Formation of crossed wall patterns in conjugate double slip; top: total dislocation density at different shear strains, bottom: dislocation density tensor component $\alpha_{31}\approx -\alpha_{32}$, all other components are zero.}
    \label{fig:walls2s}
\end{figure}

Next, we envisage a situation where two conjugate slip systems are active, which we take to be slip systems 1 and 2 in the above notation. Again we start from wall-like perturbations. If the initial perturbation has the same structure as in the single slip case, the resulting pattern is identical to the one in \figref{fig:walls1s}. If a perturbation is also present perpendicular to the slip plane of the second slip system, however, two arrays of crossed walls are forming, giving rise to a 2D cellular pattern similar to the 2D patterns investigated in Ref. \cite{wu2021cell} for a 2D model, see \figref{fig:walls2s}.
Comparison with the results of previous work clearly indicates that, for quasi two dimensional patterns imposed by 2D initial conditions, the differences between the present model and previous work are minor. This is in line with the observation that the present model contains the case of 2D dislocation systems, studied extensively in past work, as a limit case. 

To study three dimensional patterns, we consider symmetrical multiple slip where we assume initially homogeneous slip activity on all 4 slip systems (all dislocation density vectors $\Brho^i = 0$, and initial dislocation densities $\rho^i = \rho_0 + \delta \rho_i(\Br)$. For the perturbations required to initiate the patterning process, we now consider not localized planar perturbations but a generic white noise. Thus, we take the $\delta \rho^i$ to be Gaussian random fields with the correlation properties
\begin{equation}
\langle \delta \rho^i(\Br) \delta \rho^j(\Br')\rangle =
\rho_0^{2} \epsilon^2 \delta_{ij} \xi^3 \delta(\Br-\Br'),
\end{equation}
where the correlation length $\xi = \rho_0^{-1/2}$ is taken equal to the dislocation spacing. Technically, the random fields are implemented by assigning to each grid element $I$ with volume $\rho_0^{-3/2}$ a Gaussian distributed perturbation $\delta \rho(\Br_I) = \epsilon\rho_0 G_I$ where the $G_I$ are independent, standard normal distributed random variables. 

\begin{figure}[htpb]  \centering
    \includegraphics[width = \textwidth]{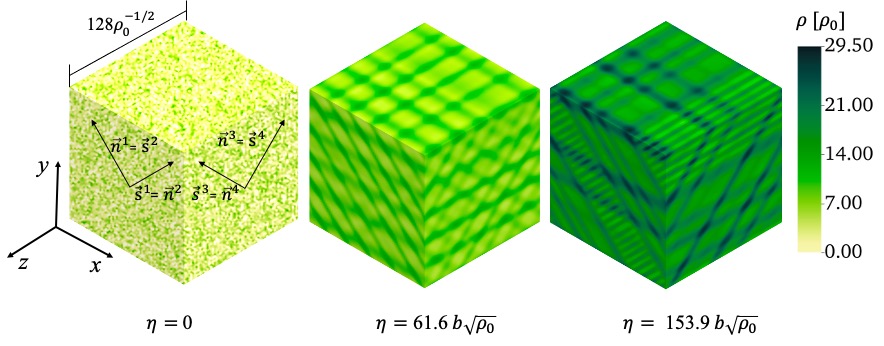}
    \includegraphics[width = \textwidth]{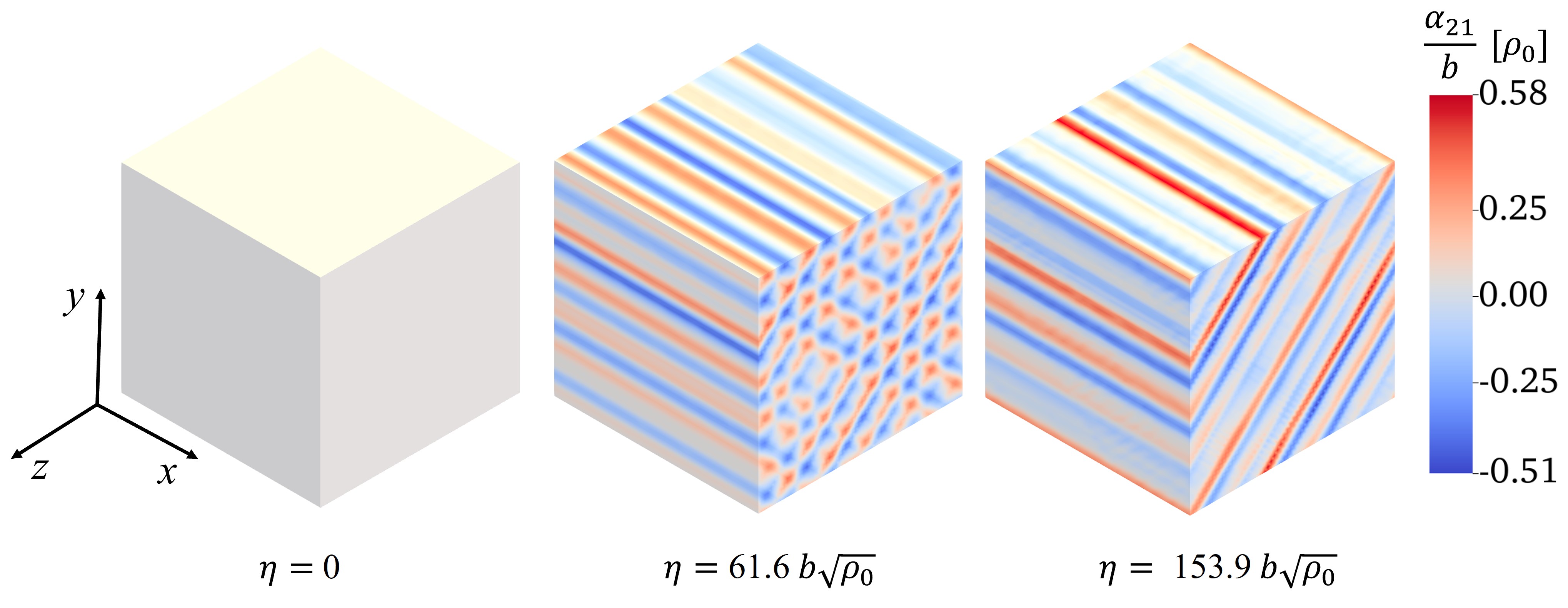}
    \includegraphics[width = \textwidth]{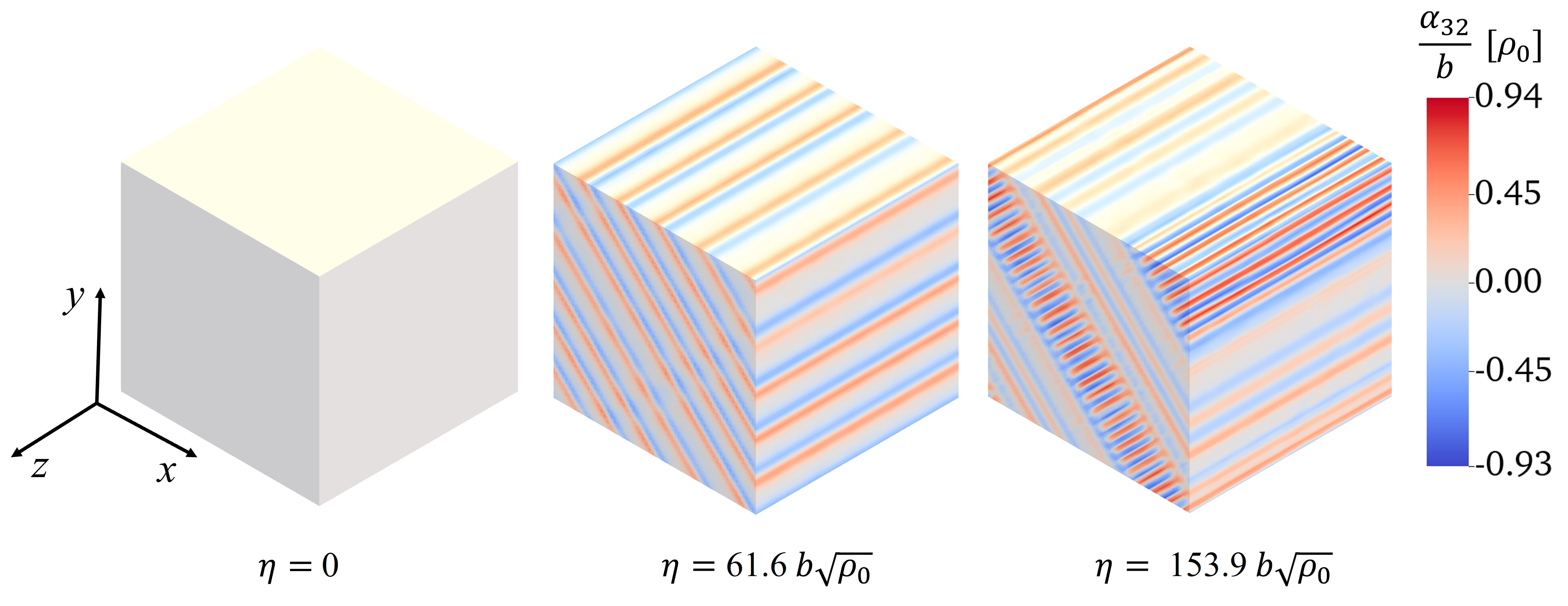}
    \caption{Formation of cellular patterns in symmetrical multiple slip; top: total dislocation density at different shear strains, bottom: dislocation density tensor components; the pattern of $\alpha_{23}$ resembles that of $\alpha_{21}$ and the pattern of $\alpha_{31}$ resembles that of $\alpha_{32}$.}
    \label{fig:cells4s}
\end{figure}
Development of the ensuing patterns is illustrated in \figref{fig:cells4s}. The initial pattern formation process up to a slip system strain of about 15$b\sqrt{\rho}$ is characterized by approximately symmetric activity of all 4 slip systems and one can discern crossed wall patterns which lead to a globally cellular structure (central graphs in \figref{fig:cells4s}). Until this point, one might think that the process simply amounts to the formation of independent wall patterns akin to those seen in \figref{fig:walls1s} on all 4 slip systems, without appreciable interactions among the different slip processes. These walls carry near zero net Burgers vector, i.e., in the terminology of Hughes et. al. \cite{hughes2003geometrically} we may speak of incidental dislocation boundaries. 

\begin{figure}[htpb]  \centering
    \includegraphics[width = \textwidth]{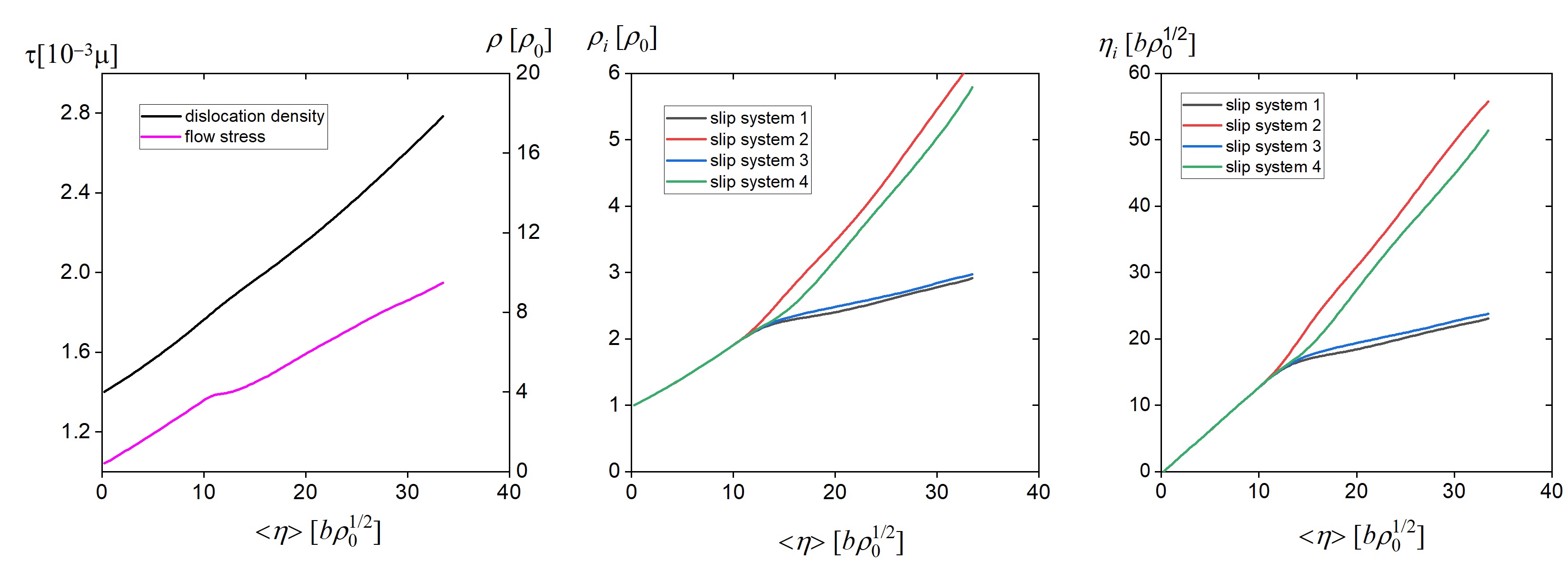}
    \caption{Evolution of stress, slip system strains, and dislocation densities during deformation in symmetrical multiple slip; left: total dislocation density and resolved shear stress on the 4 slip systems; center: shear strains on the 4 active slip systems; right: dislocation densities on the active slip systems; all variables are given as functions of the average shear strain $\langle \gamma \rangle = M \epsilon_{xx}$ where $M=2$.}
    \label{fig:multislip}
\end{figure}

However, in a second stage, nonlinear interactions between the patterns lead to a symmetry breaking bifurcation with the consequence that slip activity on two slip systems drops sharply while the other two systems (one of each conjugate pair) develop enhanced activity as seen in \figref{fig:multislip}. The dislocation density on these systems increases at an enhanced rate and the dislocation pattern transforms to a 'Harris tweed' like pattern (right column in \figref{fig:cells4s}) while, in the plane perpendicular to the tensile axis, the morphology of a roughly regular pattern of cells is maintained. We note that the selection of the systems with enhanced and reduced activity depends on details of the initial noise; in multiple simulations, all four symmetry equivalent combinations are observed with approximately equal probabilities. In a larger scale simulation, it is likely that the locally dominant slip systems may vary between different locations, leading to the emergence of geometrically necessary dislocation boundaries \cite{hughes2003geometrically}.
\section{Discussion and conclusions}
Our investigation has demonstrated that a phase-field like model of dislocation density dynamics can capture important features of dislocation patterning in three dimensions. The fundamental patterning mechanism observed in our simulations is the formation of a sequence of dislocation walls which, as argued previously \cite{wu2018instability}, can be envisaged in terms of an analogy with instabilities in traffic flow that occur if a flux variable decreases upon an increase in density of the flowing objects. At the same time, energetic considerations play an important role as the unstable patterning modes are such as to reduce long-range internal stresses as much as possible. As a consequence the emergent patterns are, on the scale of the pattern wavelength, almost internal-stress free. 

\begin{figure}[bht]
	\centering
	\hbox{}
	\includegraphics[width=0.45\textwidth]{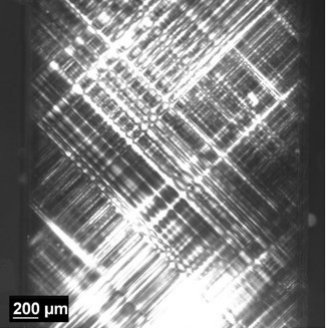}\hfill
	\includegraphics[width=0.46\textwidth]{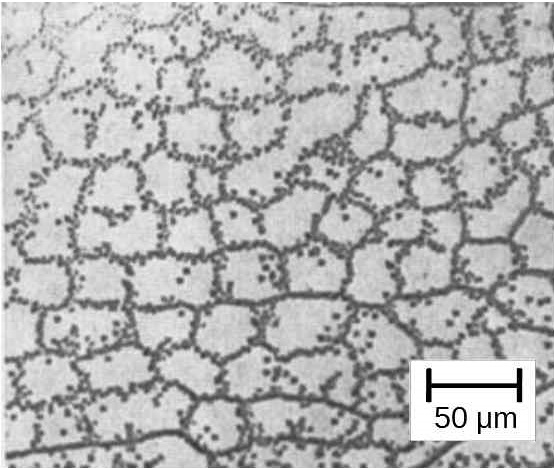}\hfill
		\caption{\label{fig:LiF}
		Cell structures in LiF; left: birefringerence image of the (001) surface of a $(100)$ oriented single crystal
		showing slip activity on the orthogonal $(1\bar{1}0)[110]$ and $(110)[1\bar{1}0]$ slip systems;
		right: etch pit pattern on a (100) cross section after deformation under a creep load of $\sigma = 5.9$ MPa ($\tau = 2.45$ MPa) to a creep strain of $\epsilon^p = 0.05$  ($\gamma = 0.1$), deformation temperature 773K, averaged dislocation density $\rho = 3 \times 10^{11}$ m$^{-2}$ \cite{streb1973steady}.}
\end{figure}

The observed pattern morphology closely matches actual observations in crystals with B1 lattice structure as shown in \figref{fig:LiF}. In a cross section perpendicular to the tensile axis, the pattern is composed of approximately square cells (\figref{fig:LiF},right) whereas on surfaces parallel to the tensile axis, we observe patterns of intersecting walls where the dominant slip system exhibits spatial variations on scales of about 10 wall spacings (\figref{fig:LiF},left). 
In summary, our investigation has demonstrated a systematic strategy for formulating energetically consistent evolution equations for dislocation densites, based upon a combination of kinematic equations for density evolution with a dislocation density dependent energy functional. Formation of dislocation patterns is a natural corollary of this formulation. We have here focused on a geometrically very simple situation, namely that of B1 lattice structures. Extending the formalism to more common lattice types, and including more sophisticated evolution equations for curvature, remains an important challenge for future work.

\section*{Acknowledgement}
We gratefully acknowledge joint support by DFG and CSC under grant no. M-5011. M.Z and Y.Z also acknowledge support by DFG under grant no Za171/13-1.
 
\section*{References}
\bibliographystyle{iopart-num}
\bibliography{phasefield.bib}

\end{document}